# Tuning Electric Polarization via Exchange Striction Interaction in $CaMn_7O_{12}$ by Sr-Doping

A. Nonato[1,*], S. Yáñez-Vilar[2], J. Mira[2], M. A. Señarís-Rodríguez[3], M. Sánchez Andújar[3], J. Agostinho Moreira[4], A. Almeida[4], R. X. Silva[5] and C.W.A. Paschoal[6,*]

[1]*Coordenação de Ciências Naturais, Universidade Federal do Maranhão, Campus do Bacabal, 65700-000, Bacabal - MA, Brazil.*

[2]*Departamento de Física Aplicada and Instituto de Materiais (iMATUS), Universidade de Santiago de Compostela, 15782 Santiago de Compostela, Spain*

[3]*Universidade da Coruña, QUIMOLMAT Group, Dpt. Chemistry, Faculty of Science and Centro Interdisciplinar de Química e Bioloxía (CICA), Zapateira, 15071, A Coruña, Spain.*

[4]*Departamento de Física, IFIMUP and IN-Institute of Nanoscience and Nanotechnology, Faculdade de Ciências, Universidade do Porto, Rua do Campo Alegre 687, 4169-007 Porto, Portugal.*

[5]*Centro de Ciência e Tecnologia em Energia e Sustentabilidade, Universidade Federal do Recôncavo da Bahia, Feira de Santana, Bahia 44085-132, Brazil.*

[6]*Departamento de Física, Universidade Federal do Ceará, Campus do Pici, 65455-900, Fortaleza - CE, Brazil.*

*Contact author: ariel.nonato@ufma.br
*Contact author: paschoal.william@fisica.ufc.br

**ABSTRACT**. Magnetoelectric (ME) materials displaying strong magnetically induced polarization have attracted considerable interest due to their potential applications in spintronics and various fast electrically controlled magnetic devices. CaMn$_7$O$_{12}$ (CMO) stands out for its giant spin-induced ferroelectric polarization. However, the origin of the induced electric polarization in CMO remains highly controversial and continues to be a subject of ongoing debate. In this paper through room temperature X-ray powder diffraction (XRPD), temperature-dependent magnetic susceptibility, and thermally stimulated depolarizing current (TSDC) measurements, we provide experimental evidence for a route to tune the magnetically induced polarization by modifying the exchange-striction in CMO via Sr-doping. Our findings demonstrate that the large and broad density peaks observed near the first magnetic phase transition (T$_{N1}$ ~ 90 K) indicate contributions to the TSDC density from both extrinsic thermally stimulated depolarization processes and intrinsic pyroelectric current arising from magnetically induced polarization changes. We suggest that the reduction in induced electric polarization in CMO originates from an increase in the Mn$^{3+}$–O–Mn$^{4+}$ bond angle due to Sr$^{2+}$ doping, which weakens the exchange-striction interaction. Meanwhile, the Dzyaloshinskii–Moriya (DM) effect determines the direction of the induced electric polarization. Our result sheds light on understanding the intriguing giant-induced polarization in CMO and similar compounds with complex magnetic structures.

## I. INTRODUCTION.

Magnetoelectric (ME) materials, which exhibit coupling between electric polarization and magnetic ordering, have recently attracted significant attention due to their potential applications in spintronics and electrically controlled magnetic devices [1–9]. The magnetically-induced ferroelectrics, known as type II multiferroics, are one of the most promising magnetoelectrics because of the high coupling between the electric and magnetic ferroic orders [10–17]. However, most candidates exhibit weak induced-electric polarization, which limits their potential for further applications [18,19].

In this context, CaMn$_7$O$_{12}$ (CMO) emerges as singular multiferroic material with a quadruple perovskite structure with extended formula (CaMn$_3^{3+}$)(Mn$_3^{3+}$ Mn$^{4+}$)O$_{12}$, where ¾ of the A-sites are occupied by Mn$^{3+}$ ions and ¼ by Ca$^{2+}$, while the B sites are occupied by Mn$^{3+}$ and Mn$^{4+}$ below 440 K. This material displays two antiferromagnetic AFM helical spin orderings (named AFM I and AFM II) at T$_{N1}$ = 90 K and T$_{N2}$ = 48 K [20]. Notably, in the recent past, a relatively large magnetically-induced spontaneous polarization (~2870 μCm$^{-2}$) was observed in CMO below the first antiferromagnetic (AFM I) transition [21–23]. However, since the direction of electric polarization in CMO lies perpendicular to the plane of rotation of Mn spins in the helical magnetic structure the induced electric polarization in CMO cannot be explained by the spin–current mechanism [24].

In the last decade, there has been considerable debate in the literature regarding the origin of induced-electric polarization in multiferroic CMO [24–29]. Two models have been distinguished for explaining the origin of magnetically-induced polarization in CMO: (i) the first proposes a combined effect involving exchange striction and Dzyaloshinskii-Moriya (DM) interactions [30], with exchange-striction interaction being responsible for the large induced ferroelectric polarization (**P**) in CMO [25,28,31], (ii) the second model argues that, given CMO's noncentrosymmetric crystal structure, the significant spontaneous polarization observed in the material arises from the combination of $p - d$ hybridization and exchange-striction [25,29].

Most recently, Lu *et al*. investigated the origin of the giant spin-induced electric polarization in AMn$_7$O$_{12}$ (A=Bi$_{0.5}$Ag$_{0.5}$) [26]. Their findings strongly suggest the giant electric polarization observed in the A$^{2+}$Mn$_7$O$_{12}$ perovskite family does not depend on orbital ordering. Instead, the exchange-striction model is the most plausible explanation for the ME coupling mechanism in these materials. This new evidence strongly contradicts the idea previously proposed by Perks *et al*., which suggested that orbital ordering plays a crucial role in stabilizing the chiral magnetic structure in CMO. This points out that the predominant mechanism for inducing electric polarization in CMO is more likely due to exchange-striction interaction, where the spontaneous polarization is strongly coupled to the helical magnetic order. Otherwise, some studies have shown that the high values of electric polarization observed in CMO originate from thermally stimulated currents (TSC) [32] [33]. All these issues highlight that the mechanism behind the multiferroicity in the compound remains controversial.

Regarding this, we point out some experimental findings that suggest the electric polarization observed below T$_{N1}$ cannot purely originate from extrinsic effects: (i) a strong magnetoelectric coupling is observed just below T$_{N1}$ [34,35]; (ii) an anomaly in the dielectric constant is observed at $T_{N1}$ [35]; (iii) recently, the exchange-striction interaction at the first magnetic transition, investigated by PXRD, has been recognized as crucial for the emergence of simultaneous electric polarization [36]; (iv) a significant dependence of electric polarization on A-site cation doping was observed below $T_{N1}$ [26,34,37]: and, according to A. Nonato *et al.* the phonon anomalies observed at $T_{N1} \sim 90$ K are evidences for the to spin-phonon coupling in CMO [38].

In this paper, based on X-ray powder diffraction (XRPD), temperature-dependent magnetic susceptibility measurements, and thermally stimulated depolarizing currents analysis, we provide insights into the origin of the observed electric polarization below the first magnetic phase transition ($T_{N1}$). For the first time, we provide evidence of a route to tuning electric polarization via both the inverse Dzyaloshinskii-Moriya (DM) effect and exchange-striction interaction, achieved by modifying the $Mn^{3+}$-O-$Mn^{4+}$ bond angle through Sr-doping in CMO. This is achieved by doping the A-site of CMO with small concentrations of $Sr^{2+}$ (up to $\leq 0.3$), which modifies the effective ionic radius of the A-cations, leading to variations in the Mn-O-Mn bond angles. Our results reveal that the broad peaks observed in Sr-doped CMO samples around $T_{N1} \approx 90$ K are attributable to thermally stimulated depolarization currents (TSDC), with contributions from both extrinsic thermally stimulated depolarization processes and intrinsic pyroelectric currents. Additionally, our approach provides evidence that the reduction in intrinsic polarization observed below 90 K in Sr-doped CMO primarily originates from the weakening exchange-striction strength. The dependence of magnetically induced electric polarization on Sr-concentration in CMO is discussed. This study opens new horizons in understanding the mechanisms governing magnetically induced electric polarization in CMO and other complex magnetic structures.

## II. EXPERIMENTAL DETAILS

Sr-doped CMO polycrystalline samples, $Ca_{1-x}Sr_xMn_7O_{12}$ ($0 \leq x \leq 0.30$) were synthesized by the Pechini method [39] following the procedure described for the undoped compound [40,41], using $SrCO_3$ (Panreac, >96%), $CaCO_3$ (Panreac, >98.6%) and $Mn(NO_3) \cdot 2H_2O$ (Aldrich, >98%) as starting reagents. The resin initially obtained was decomposed at 400°C, forming the precursor powders. The resulting precursor powders were grounded and then heated in air at 800 °C/60 h, 900 °C/24 h, and 950 °C/24 h, with intermediate grinding and pelletizing. The resultant powders were pressed into pellets and sintered in air at 970 °C for 60 h.

The obtained samples were characterized by X-ray powder diffraction (XRPD) at room temperature using a Siemens D-5000 diffractometer with $Cu(K_\alpha)$ radiation. The obtained XRPD data were analyzed by the Rietveld profile analysis using the GSAS [42,43]. Their magnetic properties were studied in a Quantum Design MPMS Squid magnetometer, in which zero-field-cooled (ZFC) and field-cooled (FC) magnetic susceptibility data were recorded under a field of 1000 Oe in the temperature range of $5 \leq T(K) \leq 300$. The thermally stimulated depolarization current (TSDC) of the samples was measured using a Keithley 617 electrometer with the samples enclosed in a helium-closed cycle refrigerator. The samples were prepared as pellets with a pair of electrodes attached to the two flat surfaces of the sample, which had been sputter-coated with gold. Thermally stimulated depolarizing current (TSDC) measurements were performed during the heating process at constant rates of 4 K/min, 6 K/min, and 8 K/min. The samples were first subjected to an electric field at 100 K, followed by cooling down to 14 K using the same temperature ramp intended for the heating measurement. After reaching 14 K, the samples were short-circuited for ~30 minutes to prevent free charges. Finally, the depolarizing current was recorded between 14 K and 150 K with the selected temperature ramp. In addition, the direction of polarization could be inverted by switching the electric field, resulting in symmetric induced polarization curves around the temperature axis.

## III. RESULTS AND DISCUSSION

### A. Structural analysis

Figure 1 shows the XRPD patterns obtained for the $Ca_{1-x}Sr_xMn_7O_{12}$ samples ($0 \leq x \leq 0.30$). As in the parent CMO compound, at room temperature, all samples have a distorted perovskite structure with trigonal symmetry belonging to the space group $R\bar{3}$ (Figure S1). Nevertheless, the presence of weak extra peaks reveals the presence of small traces of $Mn_3O_4$ (hausmannite). It is generally challenging to obtain this compound as a completely single-phase material, and traces of impurity have also been observed by other authors [31,44]. However, quantitative analysis of our XRPD data by the Rietveld refinement reveals that these impurities, when observed, always are at low concentrations ($Mn_3O_4 < 2\%$). Detailed structural parameters for the $Ca_{1-x}Sr_xMn_7O_{12}$ ($0 \leq x \leq 0.30$) are shown in Table SI of the supplementary material.

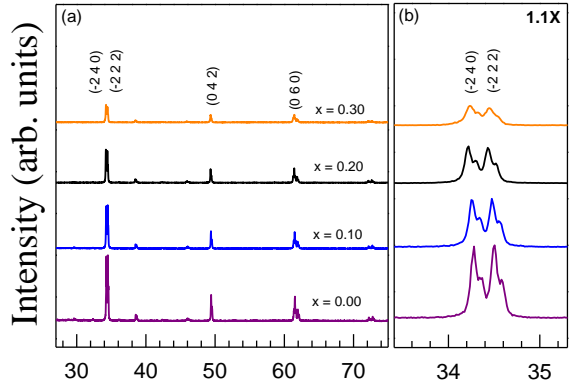

FIG 1. (a) Room temperature XRPD patterns of $Ca_{1-x}Sr_xMn_7O_{12}$ ($0.00 \leq x \leq 0.30$). (b) Detailed view of the most representative diffraction peaks evolution upon Sr-doping.

Figure 1(b) clearly shows that no structural phase transition was induced into CMO upon Sr-doping, the compounds retain their trigonal symmetry (S.G: $R\bar{3}$). Additionally, Sr-doping results in an increase of the $a$ and $c$ lattice parameters and an expansion of the cell volume (see Figure 2), as substitution of the smaller $Ca^{2+}$ by the larger $Sr^{2+}$ [45] implies an increase in the average effective ionic radius of the cations in the A site.

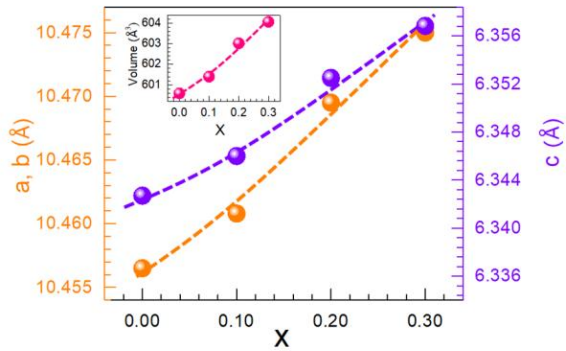

FIG 2. Variation of the lattice parameters of $Ca_{1-x}Sr_xMn_7O_{12}$ ($0.00 \leq x \leq 0.30$) as a function of Sr-doping. The inset shows the corresponding variation of the cell volume. Dashed lines are guides for the eyes.

### B. Magnetic phase transitions in Sr-doped $CaMn_7O_{12}$

Figure 3 shows the low-temperature zero-field-cooled (ZFC) and field-cooled (FC) magnetization curves of Sr-doped CMO samples, measured in an applied field of 1000 Oe. Qualitatively, the magnetic behavior of these Sr-doped samples is similar to that observed for undoped CMO compound [32,46,47]. The first magnetic phase transition ($T_{N1}$) is clearly visible in the inverse of the magnetic susceptibility $\chi^{-1}(T)$, where a subtle kink is observed at around 90 K (see inset of Figure 3a). As the temperature decreases, the magnetization abruptly increases, marking the second magnetic transition at ~ 42 K. At this temperature, a clear divergence between the ZFC and FC curves is also observed in the Sr-doped samples. It is important to highlight that this anomaly was also observed in the phonon energy at approximately 41 K in pure polycrystalline CMO [38,48]. Thus, the two magnetic transitions, $T_{N1}$ at 90 K and $T_{N2}$ at ~ 42 K, are observed and confirmed by the specific heat ($C_p$) measurements from reference [35], which exhibit distinct anomalies at both temperatures, as shown in the inset of Figure 3(a). There are no changes in the magnetic structure of Sr-doped CMO samples, suggesting their magnetic structures are rather similar. This result is in good agreement with those previously reported for $Ca_{1-x}Sr_xMn_7O_{12}$ ($0 \leq x \leq 0.50$), where neutron diffraction confirmed that the helical magnetic structure remains incommensurate at all values of x, while the fundamental magnetic wavevector increases upon Sr-substitution [49].

Additionally, another anomaly is observed in the magnetization curve at approximately $T_{N*}$ ~ 55 K. This anomaly becomes more evident when analyzing the derivative of the inverse magnetic susceptibility, as shown in Figure 3(b). Notably, this anomaly in magnetic susceptibility is more pronounced in the CMO sample with a Sr-concentration of x = 0.30, suggesting a significant change in the spin exchange interactions within the AFM II magnetic lattice due to Sr-doping. Indeed, the second AFM II transition is drastically affected by Sr doping in CMO, being observed at $T_{N2}$ = 62 K for the $SrMn_7O_{12}$ compound [50]. Additionally, this anomaly is also observed in the real part of the dielectric constant below 60 K, highlighting the interplay between dielectric properties and spin configuration (see Figure S2).

In the temperature range of 150 to 300 K (paramagnetic phase), the magnetic susceptibility of the Sr-doped samples can be modeled by the Curie-Weiss law ($\chi = \frac{C}{T-\theta}$, where $C$ is the Curie constant, $T$ is the absolute temperature, and $\theta$ is the Curie temperature) (see Figure S3). Such modeling enables the direct calculation of the Curie-Weiss constant, allowing us to estimate the effective magnetic moment as $\mu_{eff} = 2.83 C^{\frac{1}{2}} \mu_B$ [51]. Table 1 presents the constant parameters obtained for all Sr-doped samples, as well as the estimated effective magnetic moments. We found the effective magnetic moment of Sr-doped CMO samples to be very close to the expected value of 12.61 $\mu_B$ for six $Mn^{3+}$ ions and one $Mn^{4+}$ ion. Furthermore, the Curie-Weiss behavior and the negative θ value indicate a predominance of antiferromagnetic interactions in all Sr-doped samples.

TABLE I. The parameters of the Curie-Weiss Fits of Sr-doped CMO samples.

| Samples | Sr-content | θ (K) | C | $\mu_{eff}$ ($\mu_B/f.u$) |
|---|---|---|---|---|
| CMO | x = 0.00 | -27.31 | 20.42 | 12.79 |
| CSMO01 | x = 0.10 | -26.14 | 19.41 | 12.47 |
| CSMO02 | x = 0.20 | -26.02 | 20.02 | 12.66 |
| CSMO03 | x = 0.30 | -28.83 | 21.01 | 12.97 |

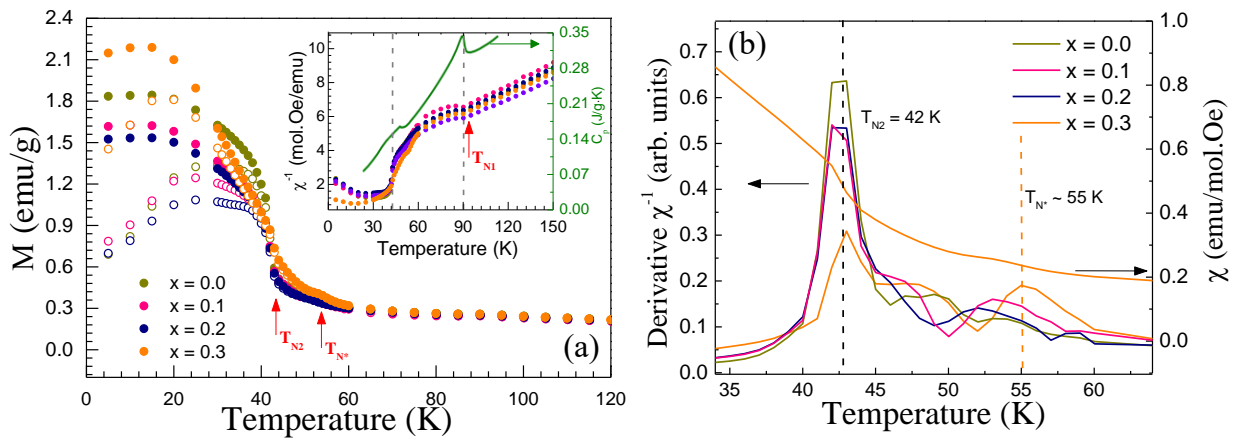

FIG 3. (a) Temperature-dependent FC (opened circles) and ZFC (filled circles) magnetization data at 1000 Oe for $Ca_{1-x}Sr_xMn_7O_{12}$ (0.00 ≤ x ≤ 0.30) (indicate **M** in the figure). The inset shows the inverse of the magnetic susceptibility (ZFC) versus temperature, with vertical dashed lines indicating $T_{N1}$ and $T_{N2}$. The specific heat ($C_P$, right axis) versus T from reference [35]. (b) Temperature-dependent derivative of the inverse magnetic susceptibility for Sr-doped CMO samples.

It is important to highlight that the CSMO03 sample exhibits a magnetic curve profile that differs from the other samples. It is observed that the difference between the ZFC and FC curves is much smaller in the CSMO03 sample. The peak of the ZFC curve occurs around 18 K, whereas in the other samples, this peak is observed around 25 K. These changes suggest that the magnetic interactions and/or domain behavior below ~42 K in the CSMO03 sample have a distinct dynamic compared to the other samples. The shift in the peak temperature of the ZFC curve and the reduced divergence between ZFC and FC curves may indicate variations in magnetic ordering or the size distribution of magnetic domains. The distinct dynamics in CSMO03 might be linked to a lower concentration of magnetic clusters or a different type of interaction between the domains, which could influence the freezing temperature or the strength of inter-domain coupling [34,52].

### C. Field-induced thermally stimulated currents (TSDC) in pure $CaMn_7O_{12}$

Figure 4(a) shows the temperature dependence of the thermally stimulated depolarizing current (TSDC) for the pure CMO compound under different field magnitudes (0.695, 3.5, 7.0, and 10.0 kV). As can be observed, for electric field values higher than 3.5 kV/cm, the peak becomes broader and exhibits a split. In this case, it is observed that the temperature where the current appears (~100 K) does not even coincide with the magnetic transition temperature ($T_{N1}$ = 90 K). Since we considered an appropriate short circuit in the Sr-doped samples, these charges originate from the orientation of ferroelectric domains during the poling process [53,54]. In this case, the observed current is not purely intrinsic to the material. In contrast, at lower fields, around 0.695 kV/cm, the shape of the pyroelectric current peak is quite similar to that reported for CMO by Zhang *et al.* [35], exhibiting a broad and asymmetric profile.

Additionally, we performed TSDC as a function of temperature at different heating rates: 4 K/min, 6 K/min, and 8 K/min, as shown in Figure 4(d). The peaks exhibit no shift along the T(K) axis, indicating that the observed current peak for the 0.695 kV/cm field is not purely extrinsic, but mostly intrinsic in origin. This behavior will be discussed in more detail later. We also observed a small anomaly in the current peak at approximately 52 K (see Figure 4b), consistent with observations made by other authors [33,55].

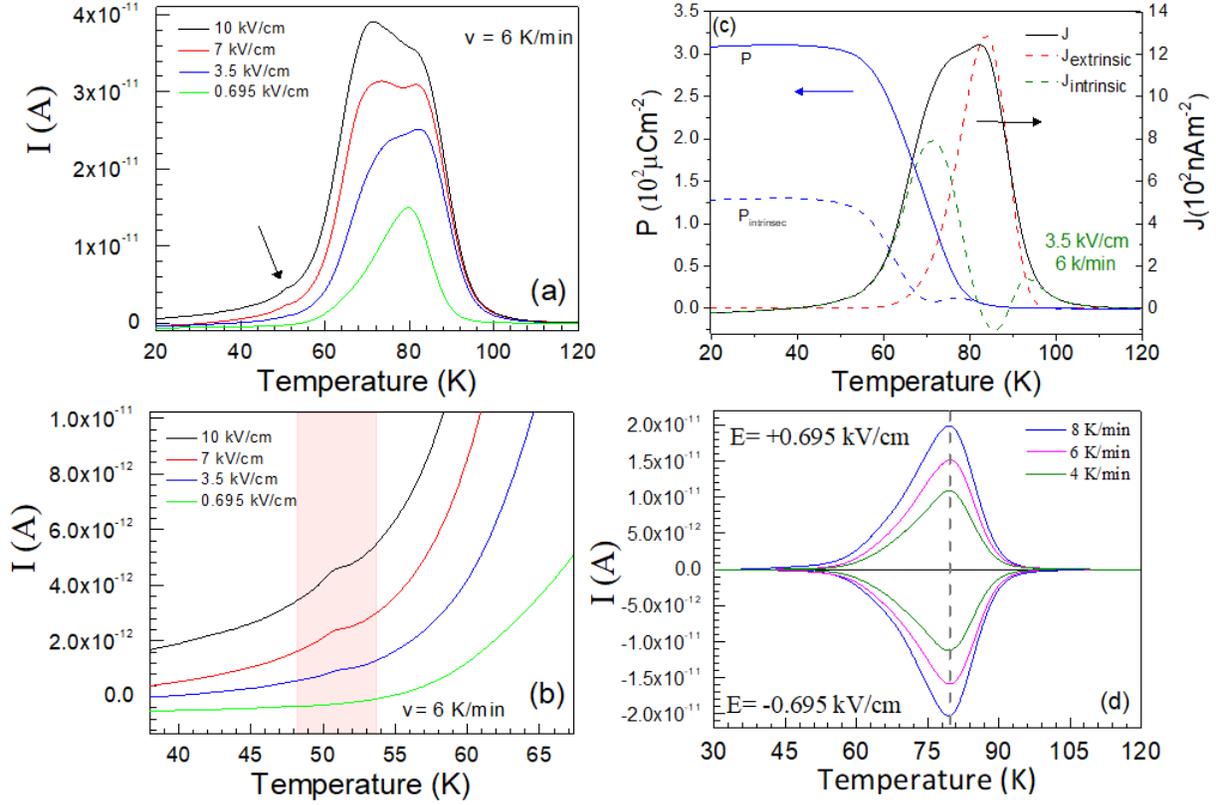

FIG 4. (a) Thermally stimulated depolarizing current (TSDC) measured at various electric field magnitudes (0.695, 3.5, 7, and 10 kV). (b) Anomaly observed in the TSDC around the transition region at $T_{N2} \sim 52$ K. (c) Temperature dependence of the TSDC density (right axis), measured with a heating rate of 6 K/min after cooling under a poling electric field of 3.50 kV/cm. The TSDC is decomposed by fitting the extrinsic component using Eq. 1 in the 80–200 K temperature range (see main text). (d) The (symmetric) pyroelectric current was measured with different heating rates (4 K/min, 6 K/min, and 8 K/min) under positive and negative poling electric fields.

The profile of the current curve for the 3.5 kV/cm field shown for pure CMO resembles what has recently been observed for BiMn$_3$Cr$_4$O$_{12}$ [56], where a thermally stimulated depolarizing current (TSDC) peak associated with extrinsic thermally stimulated depolarization processes (e.g., Maxwell-Wagner relaxation or other defect-induced relaxations [53,54]) was identified. Therefore, the TSDC density (J$_{extrinsic}$), measured between 40 K and 100 K, can be separated into two components by modeling the extrinsic peak, referred to as J$_{relaxation}$(T), using the approximated equation: [54]

$$J_{extrinsec}(T) \approx \frac{P_e}{\tau_0}\exp\left(-\frac{E}{k_B T}\right)\exp\left[-\frac{k_B T^2}{q\tau_0 E}\exp\left(-\frac{E}{k_B T}\right)\right] \quad (1)$$

where $P_e$ is a constant, $\tau_0$ is the relaxation time at infinite temperature, $E$ is the activation energy of dipolar orientation, and $q = dT/dt$ is the heating rate. Figure 4(c) displays the best fit of Eq. 1 to J(T).

To determine the intrinsic pyroelectric contribution to the TSDC density, J$_{extrinsec}$, the extrinsic component is subtracted from the total TSDC density, J. By integrating this result, we obtain the actual ferroelectric polarization, J$_{intrinsic}$(T) (see Figure 4c). We found an intrinsic polarization of ~127 µC/m², corresponding to approximately 41% of the total observed polarization (~308 µC/m²) for CMO. Hence, this model more effectively describes the magnetically induced polarization, as this intrinsic contribution arises below $T_{N1} = 90$ K.

At this point, it is interesting to discuss the significant role of the heating rate in influencing both the amplitude and the position of the peak. As the heating rate increases, the initial polarization must be released more quickly, while the dielectric responds more slowly. Thus, we would expect the peak to increase in amplitude and shift to a higher temperature. By modeling the current function for different heating rates in Figure 4(d) using Equation (1), we found that the expected shift to a higher temperature did not occur (see Figure S4). This indicates that the peak observed for the field of 0.695 kV/cm is likely intrinsic rather than extrinsic.

### D. Magnetically induced polarization in Sr-doped $CaMn_7O_{12}$

In order to check the presence of intrinsic magnetically induced ferroelectric polarization in Sr-doped CMO samples, we employed a field-induced thermally stimulated depolarizing currents procedure. Figure 5(a) shows the temperature-dependent current density ($J_p$) obtained for the Sr-doped samples, measured under a $q = 8$K/min heating rate. In contrast, Figure 5(b) shows the polarization (**P**) obtained by integrating the current density ($J_p$). The pyroelectric current peaks observed for all Sr-doped CMO samples exhibit characteristics like those previously reported for pure $CaMn_7O_{12}$ [35]. We observed a broad peak below ~ 90 K in all samples, although it does not occur exactly at 90 K for every sample (see Figure S5). In fact, in many type II - multiferroic materials, where polarization is induced by a specific magnetic structure, the electric current peak often appears broader, as observed in $LaMn_3Cr_4O_{12}$ [57], $Y_2MnCrO_6$ [58], and e $NdCrTiO_5$ [59]. As shown in Figure S5, we could tune and reverse the electric current values by changing the sign of the poling electric field in all Sr-doped CMO samples, and the absolute values of the current are nearly proportional to the temperature sweeping rate. However, distinct characteristics were observed in Sr-doped CMO samples, where the peak shifts occur at different heating rates, and some curves exhibit double or even triple peaks, which indicates the presence of thermally stimulated depolarization currents (TSDC).

In Figure 5(c), we observed a linear reduction in CMO total polarization (**P**) with increasing Sr-doping, where the sample with x = 0.30 exhibits only a quarter of the polarization observed in the undoped CMO compound. The reduction of polarization in Sr-doped CMO was also observed by Jain *et al*. at a Sr-concentration of x = 0.10 [34]. In this paper, we observed a polarization of ~134 $\mu C/m^2$ for E = 1.0 kV/cm, value which is in agreement with that observed by Zhang *et al.* [60] in a polycrystalline CMO sample (240 $\mu C/m^2$ for E = 3.5 kV/cm) when considering the same field magnitude. Of course, this polarization value is lower compared to that observed in single crystals (2870 $\mu C/m^2$ for E = 4.4 kV/cm) [21]. The reduction in polarization observed in polycrystalline samples can be explained by the voltage-divider effect at grain boundaries, where the "effective" poling fields in ferroelectric domains are smaller than the nominal ones. Consequently, the effective poling field is greater in single crystals than in polycrystals [61].

Therefore, all these features indicate that the total polarization observed in Sr-doped CMO samples originates from both extrinsic (originated from TSDC) and intrinsic (magnetically induced electric polarization) contributions. By applying equation (1) to the observed current density across all Sr-doped CMO samples, we can effectively model both contributions to the total polarization (see Figure S7). It can be observed that the samples with x = 0.0, 0.10, 0.20, and 0.30 Sr-content can be modeled using a single peak described by equation (1). All samples display current depolarization peaks with activation energies ranging from 0.14 to 0.15 eV, and relaxation times of the processes on the order of ~$10^{-9}$ s, in perfect agreement with that recently reported for $BiMn_7O_{12}$ [62]. Additionally, Figure 5(c) shows that the intrinsic contribution ($J_{intrinsic}$) decreases approximately linearly with increasing Sr-concentration. This polarization is associated with the non-collinear spin ordering that occurs below 90 K, and we demonstrate a strong correlation between this intrinsic polarization and the variation of the angle Φ between $Mn^{3+} – O – Mn^{4+}$. As this angle increases, the induced polarization exhibits linear decay.

At this point, it is particularly important to discuss the mechanisms underlying the observed intrinsic electric polarization. Since the magnitude of electric polarization arises from the exchange-striction interaction between $Mn^{3+}$ and $Mn^{4+}$ ions at the B-site, we can gain insights into the exchange-striction mechanism in CMO by analyzing structural changes with Sr-doping. Our results reveal that Sr-doping significantly influences this interaction, which depends on the angle between Mn neighboring spins and the distance between these ions [63]. As shown in Figure 5(c), Sr-doping increases the $Mn^{3+} – O – Mn^{4+}$ bond angle, resulting in a drastic reduction in the induced electric polarization. The reduction in electric polarization indicates the Sr-induced lattice expansion modifies the exchange interactions by either increasing the Mn-O bond lengths or the Mn-O-Mn bond angles, both of which are known to directly alter exchange energies in ferromagnetic and antiferromagnetic perovskite manganites [64,65]. A similar dependence was observed in the orthorhombic $Gd_{1-x}Ho_xMnO_3$ (for x ≤ 0.5) manganite, which was investigated by Zhang et al. [66], where a decrease in the Mn – O – Mn angle induces an increase in the induced polarization.

The relationship between the structural Mn-O-Mn bond angle and the spin angles in CMO can be understood by considering the CMO spin configuration for $T_{N1}$= 90 K, as determined by neutron diffraction [12, 13]. As shown in Figure 6, in the B-site of the CMO trigonal structure, the $Mn^{3+}$ ions form triangular layers above and below the $Mn^{4+}$ ions and the oxygen ions are displaced along the direction $\mathbf{e_{12}} \times (\mathbf{S_1} \times \mathbf{S_2})$ due to Dzyaloshinskii-Moriya (DM) interactions. Such displacement implies that the $Mn^{3+} – O – Mn^{4+}$ bonds with the upper layer fare later while those with the lower layer get more acute. As a consequence of such a pattern of displacements the central $Mn^{4+}$ ions move upwards along the c-axis generating polarization [67]. Thus, the DM interaction plays a crucial role in determining the direction of the induced electric polarization [68]. However, at temperatures just below $T_{N1}$ = 90 K, due to the magnetoelectric coupling, the magnetic ordering induces lattice deformations through the exchange-striction interaction ($\mathbf{S_1} \cdot \mathbf{S_2}$) which coupled with $Mn^{3+}/Mn^{4+}$ charge order, which determines the magnitude of the observed polarization [21].

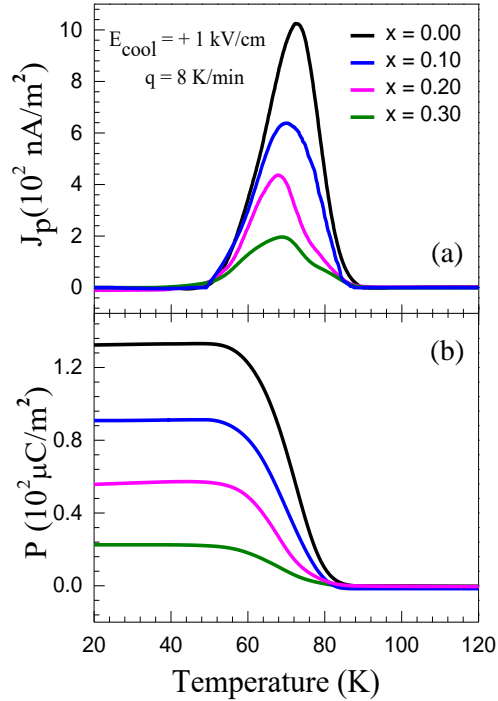
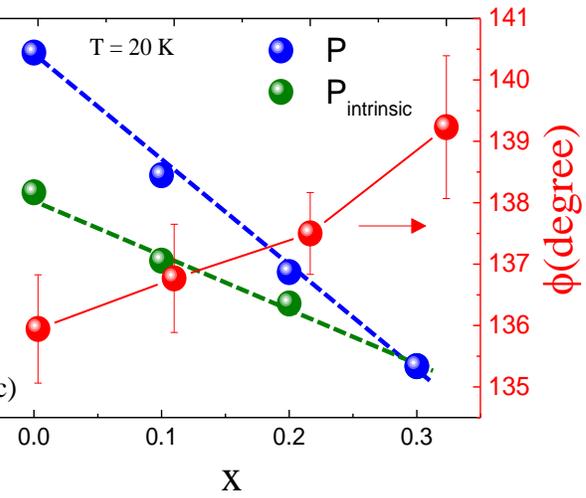

FIG 5. Temperature dependence of (a) the current density ($J_p$) and (b) the induced electric polarization (**P**) for $Ca_{1-x}Sr_xMn_7O_{12}$ (x = 0.00, 0.10, 0.20 and 0.30) samples. (c) Dependence of **P** and intrinsic (**P**$_{intrinsic}$) on the Sr-content (x), along with the variation of the angle Φ between $Mn^{3+} – O – Mn^{4+}$.

In the present study, Sr-doping increases such $Mn^{3+} – O – Mn^{4+}$ angle, as shown in Figure 5(c). Thus, when the sample reaches $T_{N1}$ under cooling, the larger angle implies a smaller $Mn^{4+}$ displacement, which results in a lower polarization. Therefore, the $Mn^{3+} – O – Mn^{4+}$ bond angle can be tuned to control the polarization in CMO. Then it is expected that larger polarization values can be induced by doping CMO in such a way that the $Mn^{3+} – O – Mn^{4+}$ bond angles get smaller than in undoped CMO. Conversely, for samples whose $Mn^{3+} – O – Mn^{4+}$ bond angle has increased up to 180°, the induced polarization should be null. It is important to stress that our results show the way to control the CMO polarization, as well as give sound experimental evidence that Johnson's model [67] fully describes the magnetically-induced electric polarization in CMO since the relationship between the induced electric polarization in CMO and the $Mn^{3+} - O - Mn^{4+}$ bond angle is defined by the ferroaxial coupling between the magnetic helical structure and the structural global rotation. Finally, based on our results, this model evidences that for large $Mn^{3+} - O - Mn^{4+}$ bond angles the induced electric polarization is low, which plausibly explains why $SrMn_7O_{12}$ has weak or no ferroelectricity [69].

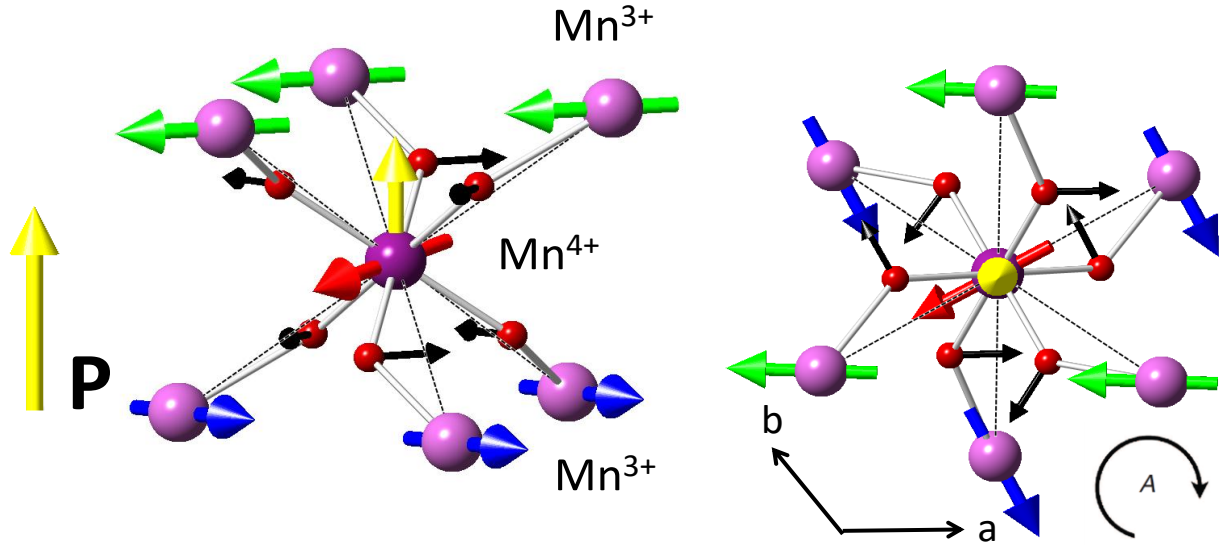

FIG 6. Schematic representation of the $CaMn_7O_{12}$ magnetic structure at low temperatures in perspective (right) and a top view (left). The light purple spheres indicate $Mn^{3+}$ at B-sites, while the dark purple spheres correspond to $Mn^{4+}$ at B-sites. Black arrows indicate the direction of the oxygen displacements due to DM interaction. Green, red, and blue arrows represent the spin directions of $Mn^{3+}$ ions in the upper plane, and of the $Mn^{4+}$ and $Mn^{3+}$ ions in the lower plane, respectively. The yellow arrow signals the polarization direction due to the $Mn^{4+}$ displacement. The ferroaxial vector **A**, which is along the -3 rhombohedral axis, is indicated in the top view (left). (Same schematic representation as proposed in reference [67]).

## IV. CONCLUSIONS

The effects of low concentrations of Sr-doping on the structural, magnetic, and electric properties of polycrystalline $CaMn_7O_{12}$ (CMO) were investigated. Structural analysis reveals that all samples exhibit a distorted perovskite structure with trigonal symmetry, belonging to the space group $R\bar{3}c$. The temperature dependence of magnetic susceptibility shows anomalies at $T_{N1}=90$ K and $T_{N2} \sim 42$ K for all Sr-doped CMO samples, without indicating significant changes in the helical magnetic structure. Thermally stimulated depolarization currents (TSDC) measurements reveal large and broad peaks around the first magnetic phase transition ($T_{N1} =90$ K), which are attributed to both thermally stimulated depolarization currents (TSDC) due to extrinsic effects in Sr-doped CMO samples and intrinsic electric polarization due to the helical spin rearrangement. By modeling the extrinsic density, we estimated the intrinsic polarization contribution and evaluated its dependence on Sr-concentration in the CMO samples. In CMO, Sr-doping strongly influences the magnetically induced polarization, significantly reducing its magnitude as Sr concentration increases. We suggest that this reduction in induced electric polarization in $CaMn_7O_{12}$ originates from the increase in the $Mn^{3+} - O - Mn^{4+}$ bond angle, weakening the exchange-striction interaction. Meanwhile, the Dzyaloshinskii-Moriya (DM) effect determines the direction of the induced electric polarization. Finally, this result sheds light on understanding the intriguing giant-induced polarization in similar compounds with noncollinear magnetic structures.

## ACKNOWLEDGMENTS

The authors thank the Conselho Nacional de Desenvolvimento Científico e Tecnológico —CNPq (Grants 406322/2022-8) for financial support. The Spanish authors are grateful to Xunta de Galicia for financial support under the project GRC2014/042. The authors acknowledge Dr. Joaquim Agostinho Moreira and Dr. Abilio de Jesus Monteiro Almeida from the

Institute of Advanced Materials, Nanotechnology, and Photonics at the University of Porto for providing all experimental facilities. M.S.A. y M.A. S.R. acknowledge Grant PID2021-122532OB-I00 funded by MICIU/AEI/ 10.13039/501100011033 and by "ERDF/EU.